\documentclass[useAMS,usenatbib]{mn2e}
\usepackage{graphicx,fleqn}
\DeclareGraphicsExtensions{.eps,.ps,.eps.gz,.ps.gz,.eps.Z}
\def \apj{ApJ}
\def \apjl{ApJL}
\def \apjs{ApJ Suppl.}

\input psfig.sty

\title[{\it Swift} observations of GRB~050712.]{{\it
Swift} observations of GRB~050712}

\author[]{Massimiliano De Pasquale$^{1}$, Dirk Grupe$^{2}$,
T.S. Poole$^{1}$, A.A. Breeveld$^{1}$, \newauthor S. Zane$^{1}$,
S.R. Rosen$^{1}$, M.J. Page$^{1}$, K.O. Mason$^{1}$, D.N.
Burrows$^{2}$, H.A. Krimm$^{3,4}$, \newauthor N. Gehrels $^{3}$,
J.A. Nousek $^{2}$, P.W.A. Roming$^{2}$, S. Kobayashi$^{5}$, B. Zhang$^{6}$\\
$^{1}$ Mullard Space Science Laboratory, University College London,
Holmbury St. Mary, Dorking Surrey, RH5 6NT, UK; mdp@mssl.ucl.ac.uk \\
$^{2}$Department of Astronomy and Astrophysics, Pennsylvania State
University, 525 Davey Laboratory, University Park, PA 16802 USA \\
$^{3}$ NASA/Goddard Space Science Flight Center, Greenbelt, MD
20771, USA\\
$^{4}$Universities Space Research Association, Columbia, MD 21044,
USA \\
$^{5}$ Astrophysics Research Institute, Liverpool John Moore
University, Twelve Quays House, Birkenhead CH4 1LD \\
$^{6}$ Department of Physics, University of Nevada, Las Vegas,
NV 89154, USA}

\begin{document}

\date{Accepted...Received...}

\maketitle

\label{firstpage}

\begin{abstract}

 We present the results of X-ray and optical observations of GRB~050712
performed by {\it Swift}. The X-ray lightcurve of this burst
exhibits episodes of flares in the first 1000s, the same epoch at
which the UVOT detected an optical counterpart. A shallow X-ray
decay, with a decay slope of $\alpha=-0.73$, followed and lasted
$\sim$70~ks. This behavior can be explained in terms of activity of
the GRB ``inner engine'', with the possibility that the last flare
is caused by the interaction of the ejecta with the surrounding
medium.

We also find interesting spectral parameters for the X-ray emission.
In particular, data suggests the presence of an intrinsic absorption in
the first 1000s, which can be explained if circumburst medium clouds lie
along the line of sight.


\end{abstract}


\section{INTRODUCTION}
\label{intro}

 Follow-up observations of Gamma-Ray Bursts (GRBs) have shown that
the initial prompt emission is followed by an afterglow, i.e.
a fading X-ray, optical and radio source
that can last up to several months after the $\gamma$-ray flash.
According to the currently accepted theory, the afterglow arises
when the burst ejecta interact with the surrounding medium and
produce a shock, which propagates in the medium and heats the
electrons. The latter, cooling down through synchrotron emission,
produce the observed radiation. Studies of afterglows can thus
provide invaluable information on the central engine of GRBs, on the
circumburst medium and can possibly lead to the identification of
different subclasses in the GRB population.

 Until recently, most follow-up observations did not start until a
few hours after the GRB, when the afterglow had already faded
significantly. This situation has changed with the launch of the
{\it Swift} mission, which provides both a rapid alert of GRB
triggers to ground-based observers, and rapid (within $\sim 1$~min
after the $\gamma$-ray trigger) X-ray and optical/UV follow-up
observations of the burst afterglow. The temporal behavior of the
GRB emission at this early epoch has been thus unveiled. Several
interesting features have been discovered (see \citealt{zh05},
\citealt{no05}, \citealt{bur05b}, \citealt{rom05a} for
detailed discussions): i) the typical behavior of the X-ray emission
has been shown to consist of an initial steep decay, followed by a
shallow one about 100-1000s after the trigger; ii) a second change
of the decay slope usually occurs about 10 ks later; iii) in some
cases, a further steepening of the slope, typical signature of
collimated outflow, is observed 100-1000 ks after the trigger.
Furthermore, in a few events the initial fast decay was unobserved,
and in others X-ray flares have been detected, superimposed on  this
``template'' time behavior (e.g. \citealt{bur05b}). As for the
optical emission, it has been observed that in many cases the
optical/UV emission is much lower than expected (\citealt{rom05a}),
suggesting some cause of suppression of the flux in the
optical band.

 Even in past years, irregular temporal features have been
occasionally seen in various bursts (see \citealt{zm02b} and
references therein): just to mention some examples these include a
rebrightening in GRB~970508, GRB~021004 and step-like features in
GRB~030309. Other similar peculiarities have been observed in
several cases (e.g. GRB~970228, GRB~980326, GRB~000203C). Various
interpretations have been proposed, mainly in terms of "refreshed
shocks" (\citealt{pan98}), supernova components (\citealt{bloo99},
\citealt{rei99}, \citealt{gal00}), dust echoes (\citealt{es00}) or
microlensing (\citealt{gar00}). On the other hand, signatures
detected by {\it Swift} in the GRB lightcurve at earlier times may
provide diagnostic about the nature of the injection and eventually
probe whether the energy is released impulsively during the event or
more continuously during the immediate post-burst epoch (see e.g.
\citealt{zm02b}), eventually with different Lorentz factors of the
ejecta (\citealt{pan05}).

 In this paper, we report the properties of the {\it  Swift} \ GRB~050712,
and discuss them in the context of the current models and scenarios
of GRBs. In \S~\ref{xray}, we describe the analysis of $\gamma$ and
X-ray data, while the analysis of UVOT data is shown in \S~\ref{opt}.
The possible interpretations are examined in \S~\ref{disc} and we
finally summarize our conclusions in \S~\ref{conc}.

\section{Analysis of the $\gamma$-ray and X-ray data.}
\label{xray}

The  {\it Swift} observatory (\citealt{ge05}) carries three science
instruments: the Burst Alert Telescope (BAT; \citealt{bar05}),
which locates GRBs with 3' accuracy, the narrow field X-ray
telescope (XRT, \citealt{bur05a}) and the Ultra-Violet Optical
Telescope (UVOT, \citealt{rom05a}). When BAT detects a GRB trigger,
{\it Swift} slews towards the source position within a few tens of
seconds.

\begin{figure*}
\includegraphics[angle=00,scale=0.6]{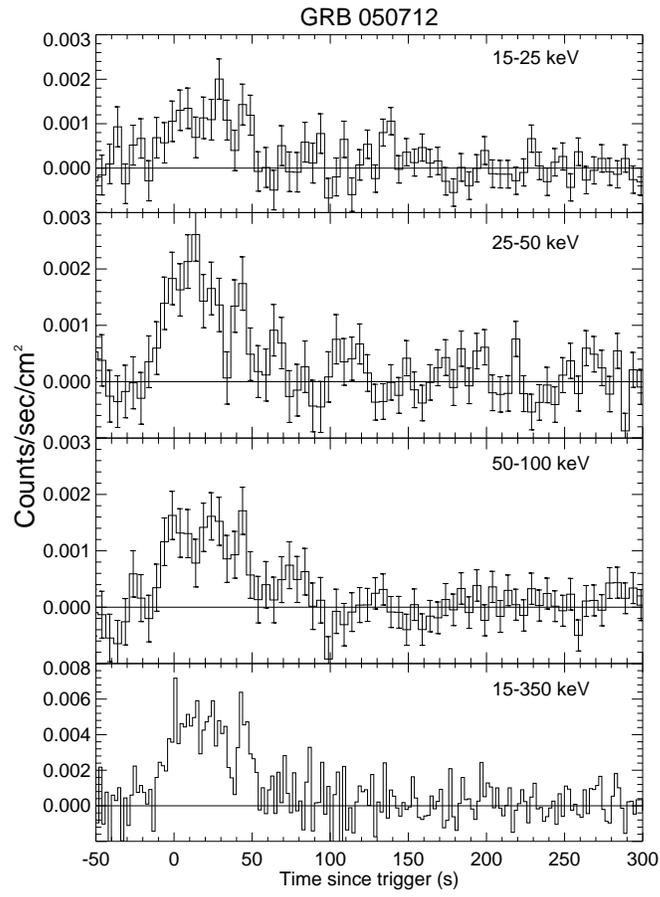}
\caption{GRB~050712 BAT lightcurve.} \label{f1}
\end{figure*}

 GRB~050712 triggered the BAT instrument at 14:00:28 UT on July 12,
2005 (\citealt{gr05a}). The refined BAT position is RA=$05^{h}
10^{m} 48^{s}$, Dec=$64^{\circ} 55' 48.2''$ with a position
uncertainty of 1.7' (90\% C.L., \citealt{mar05}, \citealt{krm05}). The
$\gamma$-ray band emission started 8s before the BAT
trigger time and the lightcurve shows a very broad peak (see
Figure~\ref{f1}), with a peak count rate of $\sim$500 counts/s
(\citealt{gr05a}).

 From the analysis of the BAT data (15-350~keV energy band) we found
that the GRB duration is t$_{90}=48$s. For the spectral analysis, we
report results obtained in the 15-150 keV band, because the mask
weighted technique has been used to subtract the background. Above
150 keV, the coded mask becomes transparent and the mask-weighting
technique is no longer effective.
 A single powerlaw provides a satisfactory fit ($\chi^2_\nu =62.8$ for 57
d.o.f.) of the BAT spectrum, with an energy index of
$\beta=-0.49\pm0.11$ (hereafter we shall assume the convention
$F_{\nu} \propto t^{\alpha} \nu^{\beta}$, where $\alpha$ is the
decay slope and $\beta$ is the spectral index; errors are reported
at 1$\sigma$ confidence level, unless specified otherwise). The
corresponding fluence is $(1.10\pm0.07)\times10^{-6}$ erg cm$^{-2}$
s$^{-1}$ (15-150~keV band).

 XRT and UVOT observations started $\sim$160s and $\sim$164s after
the trigger, respectively. Initially, the source was not bright
enough for the XRT to perform an on-board centroid
(\citealt{fa05a}). Subsequent analysis showed the presence of an
unknown X-ray source, with refined coordinates of
R.A. = $05^h$ $10^m$ $47.7^s$, Dec = 64$^{\circ}$ 54' 48.2'', with a
6'' radius error circle (90\% confidence level). This position is
within 62'' of the initial BAT position (\citealt{gr05b}). This
source subsequently faded, indicating that it was the X-ray
counterpart of GRB~050712. XRT data used in this paper were
initially taken in Windowed Timing (WT) mode (\citealt{hi05}), and
subsequently in Photon Counting (PC) mode.

    The XRT data were reduced with the {\it xrtpipeline} software,
version 0.9.9. Source and background photons of the WT mode data
were selected by {\it XSELECT} version 2.3 in a box with a length of
34.5 pixels (=$81^{''}$). The source photons of the PC mode data
were selected in a circular region with a radius of r=47$^{''}$ and
the background photons in a nearby circular region with a radius of
r=96$^{''}$. For the spectral data events with grades 0-2 and 0-12
were selected for the WT and PC mode data, respectively. For the PC
mode data of the first orbit, because of pileup, the inner 5$^{''}$
of the circle had to be discarded from the spectral analysis. The
spectral data were re-binned by {\it grppha} 3.0.0 with 20 photons
per bin. The spectra were analyzed  using XSPEC version 12.2.1. The
auxiliary response files were created with {\it xrtmkarf} and the
standard response matrices swxwt0to2\_20010101v007.rmf and
swxpc0to12\_20010101v007.rmf were used for the WT and the PC data
respectively.

The background-subtracted X-ray light curve in the 0.3-10.0 keV
energy range was constructed by using the ESO Munich Image Data
Analysis Software MIDAS (version 04Sep). The binning was dynamically
performed. At the beginning of the observations the binning was set
to 50 photons per bin while at later times it was reduced to 10
photons per bin. A pileup correction for the PC mode data of the
first orbit was applied as described in \citet{no05}. Count rates
were converted into fluxes by determining the Energy Conversion
Factors for the PC and WT mode data as described by \citet{no05}.

In order to search for spectral variability through out the
observations we also derived the temporal behaviour of the hardness
ratio. The hardness ratio has been calculated as HR = (H-S)/(H+S)
where S and H are the number of counts in the 0.3-1.0 keV  and
1.0-10.0 keV bands, respectively.

\begin{figure*}
 \includegraphics[angle=-90,scale=0.7]{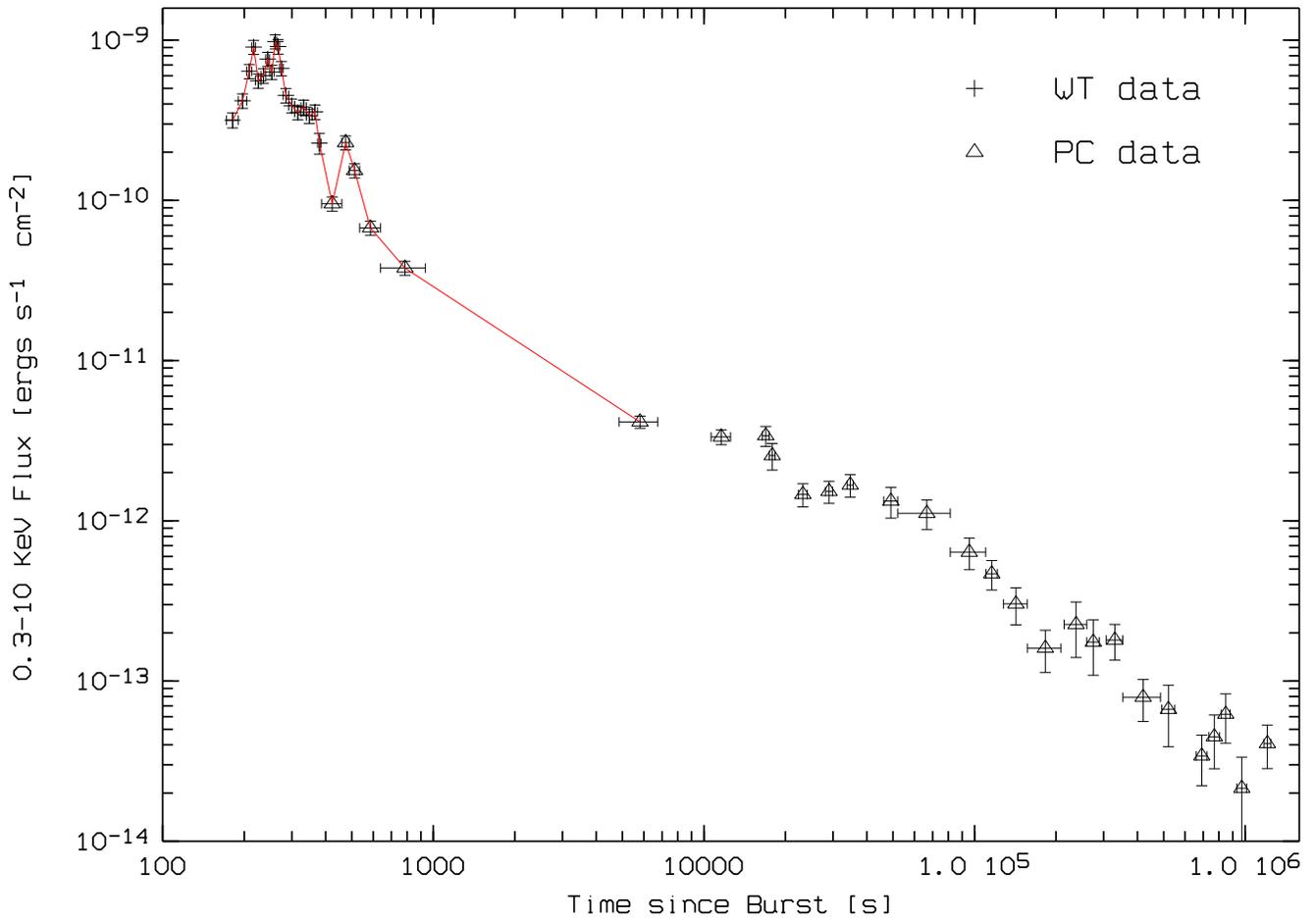}
 \caption{X-ray lightcurve of GRB~050712 detected with the {\it Swift} XRT
in the 0.3-10~keV band. Datapoints of the first 1000 sec are
connected to show the flares at 210, 280 and 480 seconds more
clearly. } \label{f2}
\end{figure*}

\begin{figure*}
\includegraphics[angle=0,scale=0.7]{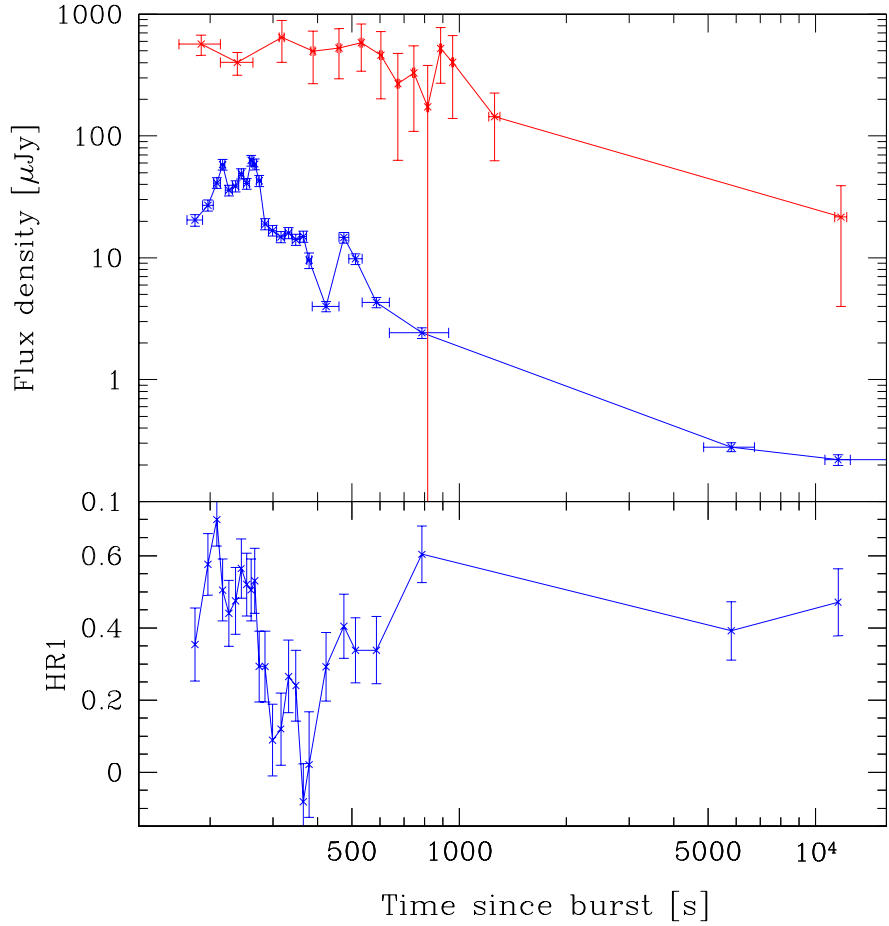}
\caption{{\it Top panel:} Zoom of the first 5000 sec {\it Swift}
GRB~050712 with XRT (0.3-10~keV, blue line) and UVOT (V filter, red
line). {\it Bottom panel:} hardness ratio evolution in the X-ray
band. The hardness ratio is defined by HR = (H-S)/(H+S) where S and
H are the counts in the 0.3-1.0 keV band and 1.0-10.0 keV band,
respectively.}
 \label{f3}
\end{figure*}

Figures~\ref{f2} and \ref{f3} show the X-ray lightcurve in the
0.3-10 keV energy band. The presence of three flares is
evident, about 210s, 240s and 480s after the BAT trigger.
We also tested if
the flares were visible in the late BAT
data. A visual inspection of the lightcurve reported in
Figure~\ref{f1} shows no obvious flare at
these epoch. Thus we looked at the data with a statistical approach.
We set up a script to go through the phase space of energy bands and
time intervals (from $t+210$s to $t+280$s) and made a set of 160 sky
images, each of which we checked for flux at the source position.
Out of the 160 trials we see 58 cases with a significance of $>2.0\sigma$,
far more than we expect from statistics. The most significant
single trial was in the interval 25-50~keV, $t+250-t+260$s,  at $4\sigma$.
However, if we look
at the whole time interval and energy band ($t+240-t+280$~s and 15-150
keV), the significance
is only 2.0 sigma ($t+240-t+280$~s in 25-50 keV gives $2.9\sigma$).
Therefore, the evidence for this peak is larger in the soft band with
respect to the total one. To make sure that there was no systematic bias,
we did a similar test using a time interval before the burst and we found
a distribution consistent with statistics.

Between the second and third flares, the lightcurve exhibits a plateau which
lasts approximately for 50s, followed by a steep decline and then by
a fast rise. After the third flare, which lasts for about
200s, there is a gap in the coverage due to the {\it
Swift} orbit, thus we cannot constrain the behaviour of the X-ray
lightcurve during this interval. However, an extrapolation of the
later X-ray lightcurve backward reconnects to the late points of the
flare, showing a shallow decline after $\sim 500$s and a possible break at
later times, i.e. around 70~ks after the trigger (see further).

 We note that the 0.3-10 keV fluence in the first 1000s (i.e. in
the interval including the three flares) is $\approx
9\times10^{-8}$ erg cm$^{-2}$, i.e.  $\sim10\%$ of the fluence
detected in the 15-150 keV band during the prompt emission phase.

\begin{figure*}
\includegraphics[angle=-90,scale=0.6]{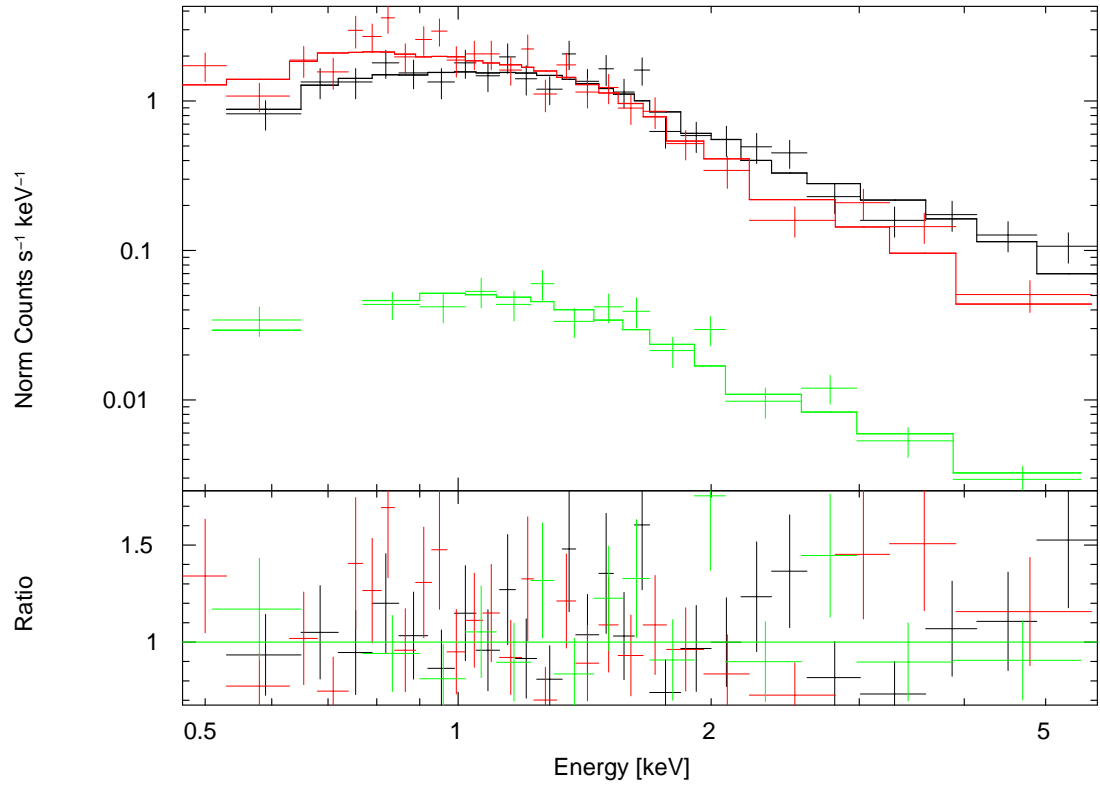}
 \caption{Evolution of the XRT spectrum during the first 1050s
from the BAT trigger. Black line: WT spectrum at the first flare
(205-281s). Red Line: WT spectrum after the first flare (281-350s).
Green line: PC data (450-1050s). Data points and best-fitting model
are shown in the top panel, while residuals are shown in the bottom
panel.} \label{f4}
\end{figure*}

 We fitted the X-ray spectra taken at different time intervals with
an absorbed powerlaw. Best fit values of column density and energy
spectral index are summarized in Table~\ref{tab_spec}, while
Fig.\ref{f3} shows the evolution of the hardness ratio.
Spectra taken before, during and after the flares are shown in
Fig.~\ref{f4}. As we can see, the X-ray data indicate a clear
spectral evolution throughout the first 1000s, with a change in the
spectral index. In particular,
after the first $\sim$200-300s the spectrum starts softening
monotonically, while, after the second peak, it changes abruptly and
the energy index increases up to a value consistent with that
observed in the late phases of the afterglow.

We also detect a photoelectric absorption in excess of the Galactic
column density, $N^{Gal}_H = 0.13\times10^{22}$cm$^{-2}$ (\cite{dl90}).
Further discussion on this issue is postponed to \S\ref{disc_3}.


\section{Optical observations by UVOT in the UV/optical.} \label{opt}

{\it Swift} UVOT detected the optical counterpart of GRB~050712 in
the V band, $\sim$200s after the BAT trigger (\citealt{rol05}). The
source was fading away, but has been positively detected by {\it
Swift} UVOT until 15000s after the trigger. The log of optical
observations is reported in Table~\ref{tab_obs}.

 The analysis of the {\it Swift} UVOT optical images was performed by
correcting for attitude drifts with an in-house software and summing
up all the exposures in the indicated time interval with
\textit{uvotimsum}, part of Swift software version
2.0\footnote{http://swift.gsfc.nasa.gov/docs/swift/analysis} .
Source counts were extracted in a circular region of 3'' radius,
while the background was extracted in a circular region of 20''
radius. In both cases, the
packages\footnote{http://heasarc.gsfc.nasa.gov/docs/software}
\textit{DS9} version 4.0 and \textit{Ximage} version 4.3.1 was used
to create the region files and extract counts, respectively. Counts
were converted in magnitudes by using the latest CALDB zero-points,
\textit{swuphot20041120v102.fits}\footnote{http://swift.gsfc.nasa.gov/docs/heasarc/caldb/data/swift/uvota}.
UVOT magnitudes can be equated to the Johnson standard.\\

 The V band lightcurve is shown in Fig.~\ref{f3}. A positive
detection has also been obtained in the U band by UVOT (see
Table~\ref{tab_obs} for values). This allows us to put an upper
limit on the redshift of GRB~050712. For redshift $z=3$, the
intervening extragalactic hydrogen would produce a U-V colour index
of $\simeq1.5$ (\citealt{lam03}, \citealt{lam00}, \citealt{zuo93})
and we would not expect an optical detection in the U band at all
for redshift larger than this value. Rather, our data suggest that
magnitudes in the V and U band are equal within errors. We can
therefore take $z=3$ as upper limit for the redshift of GRB050712.


 For completeness, we also list two successful ground-based optical
follow up observations which were triggered by the prompt {\it
Swift} localization and led to the detection of the optical
afterglow in the R band 8 hours and 14 hours after the burst trigger
(\citealt{zeh05}, \citealt{mai05}).

\section{Discussion.}
\label{disc}

\subsection{\ X-ray flares.}

\label{disc_1}

 As discussed in \S~\ref{xray}, the X-ray lightcurve of GRB~050712
exhibits at least three flares, at about 210s, 240s and
480s after the BAT trigger. These features have been observed with
{\it Swift} in several other GRBs (see \citealt{bur05b},
\citealt{fa05b}, \citealt{romano05}, \citealt{zh05}) and seem to be
characteristic of the X-ray emission detected at early times.

 In principle, the occurrence of flares can be associated with several
mechanism such as: reverse shock propagation, presence of high
density clouds surrounding the progenitor, or GRB emission
collimated in a multicomponent outflow with variable amount of
energy per solid angle, as in models involving ``structured jets''
or ``patchy shells'' (\citealt{be03}, \citealt{hu04}, \citealt{kp00}).

 However, the majority of these models, which ultimately account for
different ways in which the ejecta interact with the circumbust
medium, cannot account for the rapid onset and decline of the X-ray
flare observed in several GRBs by {\it Swift} (\citealt{zh05}).
An alternative scenario for these features is that the ``inner
engine'' of the GRBs does not switch off at the end of the main
high-energy event but it still emits shells, which continue to
collide and produce radiation. Following \citet{zm02a}, in this case the
peak energy of the flare
emission, $E_{p}$, scales as:

\begin{equation}
    E_{p} \propto L^{1/2} \Gamma^{-2} \delta t^{-1}
\label{eq1}
\end{equation}
where $L$ is the luminosity, $\Gamma$ is the Lorentz factor and
$\delta t$ is the variability time scale of the flare. The observed
X-ray peaks have luminosity lower than that of the main event, and
their fluctuations take place on a large time scale. Also, it is
expected that the environment tends to be cleaner so that $\Gamma$
may be larger. Therefore, late ``internal shocks'' can give rise to
(X-ray) bursts which are softer as compared with the prompt
($\gamma$-ray) emission. The high Lorentz factor of the shells
naturally explains fast variations of the flux, in terms of a pure
propagation effect.
Let us assume that the shells, moving at a dimensionless velocity
$\beta$, emit a photon at the time $\overline{t_{1}}$ at a certain
distance from the centre of the explosion $\overline{r_{1}}$ and a
second photon at the time $\overline{t_{2}}$ and position
$\overline{r_1}+dr$). The time $\overline{t}$ is measured in the Earth
observer's reference frame, and the distance
between this observer and the shell location is D. This first photon
arrives at the observer at the time $t_{1}=\overline{t_{1}} +D/c$,
while the second one arrives at a time $t_{1}=\overline{t_{2}} +D/c
-\beta d\overline{t}$ where $d\overline{t}=\overline{t_{2}} -
\overline{t_{1}} \simeq dr/c$. In turn, the
interval for the observer to receive the two photons is much
reduced, i.e. $dt=t_{2} -
t_{1} = (1-\beta)\overline{dt} \simeq d\overline{t}/2 \Gamma^{2} =
dr/2\Gamma^{2}c$.

In the case at hand, the 2nd flare flux (extrapolated in the 0.3-150
keV band) is $L_{F} \approx 6 \times 10^{-10}$erg/s/cm$^{2}$, the
Lorentz Factor should be as high as $\Gamma \approx 1000$ (see
~\citealt{zh05} and reference therein) and the time scale of
variation is $~40$s. The corresponding quantities during the prompt
emission are $\sim 3\times10^{-8}$erg/s/cm$^{2}$, $\Gamma \approx
300$ and $\sim$10s. For such values, equation \ref{eq1} predict a
peak energy a factor $\approx300$ lower than that in the prompt
phase. A good measurement of the peak energy of the prompt emission
of this burst is unavailable, but if we assume a standard value of
$~250$ keV (Preece et al. 2000), we can infer that the peak energy
of the 2nd flare must be at $\approx 1$ keV, i.e. in the XRT band.

During the decay of flares produced by late ``internal shocks'',
the relationship between the spectral and decay indices is given by
(\citealt{kp00}):

\begin{equation}
\label{eq2}
  \alpha=\beta -2
\end{equation}
provided that the decay slope is computed by re-setting the
time-zero point t$_{0}$ at the time of the peak (see \citealt{zh05}
and references therein). We note that several {\it Swift} bursts
exhibit a rapid decay during the early phase of the observed X-ray
emission (first hundreds of seconds after the trigger), with
spectral slope $\sim -1$ and decays slope $\sim-3$ respectively, in
agreement with the predictions of eq.~\ref{eq2}.

In the case of GRB~050712, we found that the X-ray spectral
evolution of the emission in the first 400s is in general agreement
with this ``late internal shocks'' scenario. During this phase the
X-ray spectrum of GRB~050712 softens, with the energy index
decreasing from -1.1 to -1.7. This is reminiscent of the behavior of
the prompt emission. For comparison, the spectral index of the low
energy part of the $\gamma$-ray emission detected by BAT is around
-0.5. Assuming that the first two flares have a similar
$t_{0}=180$s, we get $\alpha=-2.5\pm0.6$ and $\alpha=-3.6\pm0.77$
for their decay slopes. We also have $\beta \sim -1.7$ (see
Tab.~\ref{tab_spec}) for the spectral index of the first and second
peak. So our results are in agreement with eq.~\ref{eq2}. Also, we
may speculate that the decay phase observed around $\sim$400s after
the trigger might follow a further release of energy which is not
observed as a ``flare'' since it is superimposed on that of the
first peak. In this case, it would be difficult to fix $t_{0}$
properly. However, we find that taking $t_{0}=240$s gives a decay
slope of $\alpha= -3.4 \pm 0.4$, which again satisfies eq.~\ref{eq2}
(since during this phase the spectral index is the
same as before, i.e. $\beta \sim -1.7$).\\


 However, the situation may differ as far as the third X-ray flare,
observed at $\sim 480$s, is concerned. First, if this peak were due
to a further internal shock occurring in the fireball, then it would
be followed by a fast decay. The rise of this peak is clearly
visible at $t=440$s after the trigger. Assuming this value for
$t_{0}$ gives a decay slope of $\alpha= -0.86\pm0.02$ which, combined
with an energy spectral index of $\beta \sim -1.16$, does not satisfy
eq.~\ref{eq2}. The situation does not improve by choosing slightly
different values of $t_{0}$. Second, the spectrum of the second peak
differs ($\sim 4-5 \sigma$) from that of the previous phase, which
could suggest a different origin, while it is much more in agreement
with that observed during the late phases of the X-ray afterglow.
Late internal shocks might still be responsible for the second flare
if, for instance, the emission does not stop suddenly after this
peak and other weak flares are released superimposing on the
decaying part. In this case, the decay would be shallower than that
predicted by eq.~\ref{eq2}. Moreover, collisions of different shells
may well produce emission with diverse spectral properties, and the
rapid increase of flux in the second flare, by a factor of $\sim 3$
within $\sim 50$s, supports the internal shock scenario.

 It should be noticed that the behaviour of GRB 050712 X-ray
afterglow is rather uncommon. Flare spectra usually soften until
they reach the relatively soft afterglow. In this case, the first
flare is quite soft, but then the spectrum hardens at later times.
However, the spectrum of the last flare is similar to that of the
late afterglow, which is hard as compared with other afterglows of
other Gamma-Ray Bursts.

 An alternative possibility is that with the third flare we witness
the onset of the standard afterglow, i.e. the creation of an
``external shock'' when the fireball runs into the circumburst
medium. The standard afterglow model predicts that, as a shell of
ejecta interacts with the circumburst medium, it gives rise to a
``forward shock'', which propagates outward, and to a ``reverse
shock'', which moves inward through the ejecta (\citealt{mr97}).
During this phase, the observable characteristics depend on the
dynamical conditions of the fireball.

 Two regimes can be identified depending on the ``thickness'' of the
fireball (\citealt{sap99}, hereafter SP99). The difference between
the two cases is whether a reverse shock becomes relativistic in the
frame of unshocked material during its crossing of the shell. If
this happens, the shell is defined as ``thick''. Otherwise,
the shell is defined as ``thin''. In the case of a thin shell, SP99 have
shown that
the evolution of the afterglow is well described by a powerlaw decay if
the time is measured starting from the explosion time, which is a
good approximation of the time at which the first photons are
collected. The situation differs if the shell is ``thick'', i.e. if

 \begin{equation}
 \Delta > (E/nm_{p}c^{2})^{1/3}\gamma^{-8/3} \label{eq3}
 \end{equation}
where $\Delta$ is the initial thickness of the shell in the observer
frame, $\gamma$ and $E$ are its initial Lorentz factor and kinetic
energy, $n$ is the number of particles of the circumburst medium per
unit volume, and $m_p$ is the proton mass. If we assume typical
parameters of E$\sim 10^{52}$ erg, n =10 cm$^{-3}$ (see
\citealt{be03}) and consider that $\Delta=cT$ , where T is the burst
duration, we need to have $\gamma>70$, which is easy to satisfy
(see, however, \citealt{zhm04}). Most of the energy is released into
the surrounding material only when the shell has been crossed by the
reverse shock. Following again SP99 (see also \citealt{pi05}), this
happens after a time $t_{\Delta} \sim \Delta/c $, roughly similar to
the burst duration.
  Only from the time $t_{\Delta}$ onward, the deceleration of the shell scales
as $\gamma^{-3/8}$, in a self-similar manner, and the afterglow
decay can be fitted by a power law (or broken power law) model.

 Quite interestingly, we find that this might the case of GRB~050712.
By rescaling the zero-time to $t_{\Delta}=440$s (i.e. to the onset
of the third peak) we find that a broken powerlaw model
provides an acceptable fit of the lightcurve at $t\geq t_\Delta$,
with $\chi^{2}_{\nu} = 33.7$ for 23 degrees of freedom (see
Fig.~\ref{f7}), which is acceptable at $7\%$ C.L. The best fit
parameters are $\alpha_{1}=-0.73\pm0.01$,
$\alpha_{2}=-1.23^{-0.10}_{+0.07}$
for the two decay slopes before and after the break, and
t$_{break}=66240^{+25000} _{-10300}$s. The presence of a break is
statistically significant since a fit with a single powerlaw gives
$\chi^{2}_{\nu} = 93$ for 25 d.o.f.

 The comparison between the optical and X-ray lightcurve is also
illuminating. First of all, in the optical we do not see the strong
variability that is present in the X-ray lightcurve. We have tried
to fit the UVOT V datapoints with a rescaled X-ray lightcurve in the
first 10000s, and we have obtained a $\chi ^{2} = 38.5$ for 13
d.o.f. Therefore, the probability that the 2 lightcurves are
consistent is only $5\times10^{-4}$. The major differences in the
two lightcurves arise from the first part. In the V band lightcurve
we see no increase correspondent to the first and second X-ray
flares, no decrease at the time of the steep decay at $\sim400$s,
and again no increase at the time of the rise of the third flare.
Rather, the optical lightcurve in the first 500s is fully compatible
with being constant: a fit with a constant gives $\chi_{\nu}=1.6/4$.
This suggests that the origin of the optical emission is different
from that of the X-ray. A possible origin for the optical emission
would be ``reverse shocks'' crossing the ejecta inwards, as already
suggested in the case of GRB990123 and GRB021211 \cite{zhm04}.
Reverse shocks are mild-relativistic, so their emission is not
expected to be associated with very fast fluctuations of the
radiation, as in case of internal shocks. After it reaches the peak,
reverse shock emission is usually expected to fade away with a
powelaw decay $F \propto t^{-2}$, and it has been shown by several
authors (see Zhang \& Kobayashi 2005, Zhang et al. 2003, Kobayashi
et al. 2000) that, if ejecta are magnetized and the optical band is
below the synchrotron cooling frequency, the peak itself may be
broad and the slope more gentle, $F \propto t^{-1/2} $. Furthermore,
the emission from the reverse shock may well overlap that of the
forward shock component after the forward shock peak time (GRB
021211 may be a marginal such case, see Zhang et al. 2003 ). In the
case at hand, we see a decrease in the optical starting 500s after
the trigger, roughly after the peak in the X-ray. The optical-to-X
energy index $\beta_{OX}$ fluctuates until $\sim$500s after the
trigger, when it stabilizes at $\beta_{OX} = -0.8$, remaining
consistent with this value for the rest of the afterglow. These
clues suggest that reverse shocks may be responsable for the flat
emission registered in the first few hundreds seconds, while, from
$\sim$500s onwards, the optical and X-ray emission are produced by
the same mechanism, possibly due to the forward shock as previously
discussed.

\subsection{The X-ray afterglow}
\label{disc_2}

Regardless of the interpretation of the third flare, after the
second orbits (i.e. 5000s after the trigger) \textit{Swift} observed
the X-ray afterglow of GRB050712 and detected a break in the decay
slope. We find that the values of two decay indices as well as the
break time are weakly dependent on the exact start of the afterglow.

The initial shallow decay ($\alpha \approx -0.73$) and the break
could be explained either by a late, continuous energy injection
from the inner engine (\citealt{zm01}, \citealt{zm02b}), or,
alternatively, by a model in which the central engine activity is as
brief as the prompt emission itself but, at the end of the prompt
phase, the ejecta are released with different Lorentz factors
(\citealt{pan05}). The two scenarios are observationally
indistinguishable, although the second one reconciles better with
the scenario of the onset of the afterglow at the third peak,
as we shall see in the following.

In the first case, a source luminosity law of the kind $L\propto t^{q}$
is assumed, where t is the intrinsic time of the central engine (or
the observer's time after the cosmological time dilation
correction). Following (\citealt{zm01}, \citealt{zm02b}), the
spectral and decay slopes are linked through the relation

 \begin{equation}\label{eq4b}
 \alpha=(1-q/2)\beta +q+1 \, .
 \end{equation}

For the values of the indices we have before the break, we obtain
$q\sim-0.5$. After the break, we get $q\sim-1$. In the continuous
energy injection scenario, the former case indicates that the
central engine produces a substantial amount of energy, which
affects the fireball evolution. Instead, for $q\sim-1$, the energy
which is added is small compared to the energy already injected and
it does not affect the fireball evolution effectively.


 In the second scenario, the fastest shells initiate the forward
shock, decelerate, and are successively caught by the slowest
shells. The consequent addition of energy in the blast-wave
mitigates the deceleration and the afterglow decay rate. Assuming
that the mass $M$ of the ejecta follows the law

\begin{equation}
\label{eq4}
    M(>\gamma) \propto \gamma^{s} \, ,
\end{equation}
 where $\gamma$ is the Lorentz factor, the spectral and decay slope must be
linked through the relationship (\citealt{zh05}):

\begin{equation}\label{eq5}
    s=-(10+7q)/(2-q) \, ,
\end{equation}

 The degeneracy of the two models consists in the fact that a
non-vanishing $s$ index mimics the same effect of non-vanishing $q$
index, although the physical mechanisms involved are different.\\
From the values of $q$ reported above, we obtain $s=-2.6$ before the break
and $s=-1$ after the break.
Before the break time, the fast shells
are decelerated initially with the slow shells lag behind ballistically
and, when the fast ones are delerated enough, they are catch up by the
slow shells. It is such pile up that gives an efficient conversion of
energy:
a steep distribution of the shells is required
in order to have a significant energy injection into the blast wave,
with more energy carried by slow shells. On the other hand, if the Lorentz
factor distribution is flatter than $\simeq-1$, the injected energy is
much smaller than that already present in the blastwave, so that the
fireball dynamics is unaffected, and the afterglow enters the "normal
decay" phase.

\subsection{Absorption by the circumburst medium.}
\label{disc_3}

 There is a general, growing, evidence for the association of long GRBs
with the death of ultramassive stars (Galama et al. 2000,
Stanek et al. 2003, Hjorth et al. 2003, Zeh et al. 2004). Since
these objects have a short lifetime ($\sim10^{6}$~yrs), it is
expected that GRB explosions take place close to or inside the
original star forming regions, where a dense circumburst medium is
present. Thus, high level of absorption are expected, and
this is consistent with the observations in ($\sim 50\%$) of Swift
bursts (\citealt{cam05}, \citealt{gr05c}, \citealt{obr05}).

In order to verify the presence of an extragalactic absorber, we
have first fitted the X-ray data with an absorbed power law model,
by keeping the absorption fixed at the Galactic value of $N_H^{Gal}
= 0.13\times10^{22}$cm$^{-2}$ (Dickey \& Lockmann 1990). This gives
$\chi^{2} _{\nu} = 26.95/24$, $\chi^{2} _{\nu} = 38.84/26$,
$\chi^{2} _{\nu} = 13.1/16$ for the WT data taken during the first
and second flare, the WT data taken after the second flare, and the
PC data taken during the third flare, respectively. Later data have
not been included in this analysis, because their statistics is
quite poor. We then repeated the fit by adding an extra absorption
component, and fixing the redshift at  $z=1$. We obtain $\chi^{2}
_{\nu} = 21.5/23$, $\chi^{2} _{\nu} = 37.3/25$ and $\chi^{2} _{\nu}
= 12.10/15$, for the same three segments. We have calculated the
probability of a chance improvement by means of the F-test (see,
however, Protassov et al. 2002). We get $P=0.02$, $P=0.20$ and
$P=0.28$. Very similar results have been obtained by varying the
redshift from z=0.5 to z=2. Results of the fits have been reported
in Table ~\ref{tab_spec}. Therefore, data gives a marginal
indication for some intrinsic absorption for GRB~050712, at least in
the first interval. Assuming a putative redshift $z=1$, the value of
the excess column density would be $N_{H}=5.6^{+2.4} _{-2.1} \times
10^{21}$cm$^{-2}$, $N_{H}=1.88^{+1.79} _{1.15} \times
10^{21}$cm$^{-2}$ and $N_{H}=1.65^{+2.01} _{-1.00} \times
10^{21}$cm$^{-2}$: there could be some indication of a decreasing
column density, although all these values are also consistent with
$N_{H}\sim3 \times 10^{21}$cm$^{-2}$.

\section{Conclusions.}
\label{conc}

We presented {\it Swift} observations of the GRB~050712 and we
discussed the properties of its optical and X-ray afterglows. The
X-ray light curve of this burst does not decay immediately
after the high energy event, but it shows two episodes of flares in
the first $\sim1000$s. We find that the first and second flare is
likely to be explained in terms of late internal shock emission.
However, this might be not the case for the third flare.
The different interpretation is based on the fact that during the
second rise of the X-ray flux, the XRT spectrum is
different from that detected during the previous phases, and more
similar to that observed at the late afterglow epochs. Furthermore,
the flux of the ``flare'' can be connected with the
late afterglow lightcurve with a broken powerlaw model, if the zero
time is rescaled to the time of the onset of the peak. This is what
is expected if the ejecta have been crossed by the reverse
shocks, roughly at a time similar to that of the burst duration,
and the onset of the external shock follows. Moreover,
after the epoch of the third flare we observe a
steepening of the optical-to-X-ray spectral index, $\beta_{OX}$. This
would suggest an increase in the optical emission,
perhaps due to the contribution of the starting forward shock
running in the circumburst medium. $\beta_{OX}$ takes the value that
will have for the remaining afterglow from this epoch. Although an
internal shock interpretation can not be completely ruled out, all
these findings may also suggest that the last ``peak'' represent the
beginning of the standard afterglow phase.\\
The spectral fit also suggests some hints of intrisic absorption, at
least in the first $\sim 300$s. Assuming z=1, the value would be
$\sim3\times10^{21}$ cm$^{-1}$, which is in the range of values for
dense giant molecular clouds. If real, this  finding is consistent
with the idea of massive progenitors for GRBs. We cannot firmly
establish whether or not the column density decreases at later times
(e.g. by progressive ionization); we only note that, if the
intrinsic absorption is not changing, it implies that the absorbing
medium should be not very close to the place where the GRB took
place (\citealt{lap02}).

\begin{figure*}
 \includegraphics[angle=-90,scale=0.7]{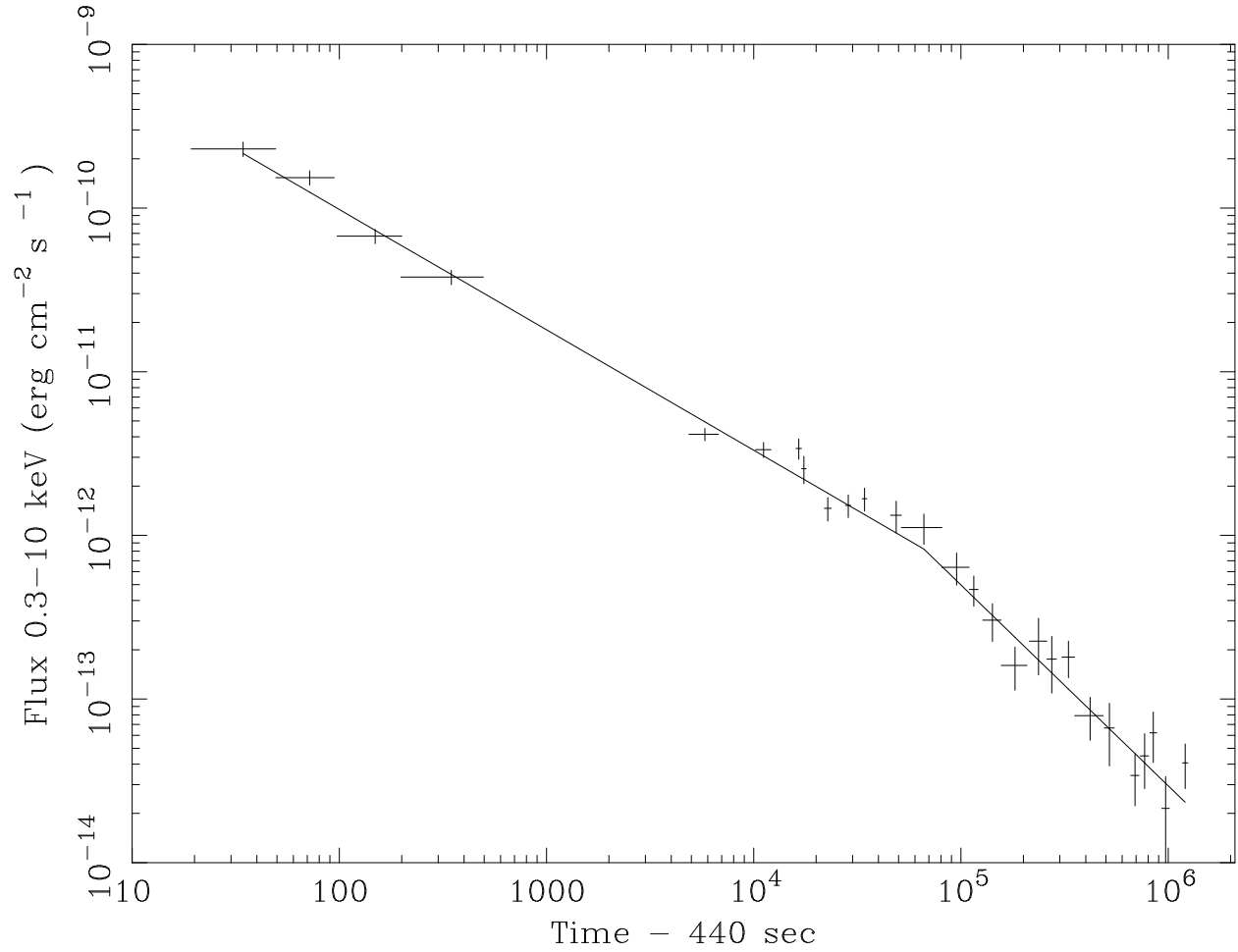}
 \caption{A broken power-law fit of the {\it Swift} XRT
lightcurve of GRB~050712, computed assuming t$_{0}=440$s (see text for
details).}
\label{f7}
\end{figure*}

The fit presented in Fig.~\ref{f7} shows that, up to $\sim 80~$ks,
the observed decay slope is rather shallow, with a slope
$\alpha_{1}\simeq-0.73$. This might be explained if a residual,
continuous energy injection from the inner engine lasted at late
times, with a luminosity law $L \propto t^{-0.5}$. A different
explanation, that does not require a late-time reactivation of the
central engine, is that ejecta are released with different Lorentz
factor according to the distribution $M(>\gamma) \propto
\gamma^{-2.6}$ (\citealt{zh05}, \citealt{pan05}). The decay slope
observed after the break, $\alpha_{2} \simeq -1.23$ is instead close
to the standard decay slope observed in X-ray afterglows $\sim1$ day
after the $\gamma$-ray event (\citealt{dp05}, \citealt{gen05},
\citealt{no05}, \citealt{zh05}).

The constraint to the redshift ($z\leq3$) enables us to set an upper
limit on the energy emitted in $\gamma$-ray and X-rays by
GRB~050712. If we assume $z=3$ (which is \textit{the} upper limit),
and a spectral shape with low energy spectral index -0.5, typical
break energy of $E_{0}=300$ keV and high energy slope = -1.5,  the
k-corrected 1-10000~keV energy emitted by GRB~050712 is
$2.9\times10^{53}$~erg.
\newline

Acknowledgements. We are grateful to an anonymous referee
for his/fer suggestions that led to a substantial improvement of the
draft. SZ also thanks PPARC for support through an Advanced
Fellowship.


\begin{table*}
\begin{center}
\begin{tabular}{ccccc|}
   \hline \\
   Section(s after trigger)  & $\beta$ & N$_{H}\times  10^{21}$  cm$^{-2}$
at
   z=1&N$_{H}\times  10^{21}$  cm$^{-2}$ at z=2 & $\chi^2_\nu$/(d.o.f.) \\
   \hline
160-281                   & $-1.10^{+0.16}  _{-0.17}$  & $5.60^{+2.40} _{-2.10} $ & $15.2 ^{+3.96} _{-3.45}$   &$21.5/23$  \\
281-350                   & $-1.71^{-0.21}  _{+0.18}$  & $1.88^{+1.79} _{-1.15} $ & $5.26 ^{+4.85} _{-3.20}$   &$37.3/25$  \\
440-1050                  & $-1.16^{+0.18}  _{-0.20}$  & $1.65^{+2.01} _{-1.00} $ & $4.34 ^{+5.39} _{-2.63}$   &$12.1/15$  \\
5000-1.2e6                & $-0.96^{+0.23}  _{-0.25}$  & $1.30^{+6.50} _{-5.85} $ & $3.88 ^{+19.40} _{-16.65}$ &$4.4/7  $  \\
   \hline
 \end{tabular}
\caption{Values of the best fit parameters for the spectral fit of
GRB~050712 X-ray afterglow. Data have been divided into 5 temporal
sections: WT data before the first flare, WT data before the first
flare, WT data after the first flare, PC data at the second flare,
and PC data taken during the late afterglow. The column
density reported in the table is the intrinsic one, calculated for two
different values of the
redshift,  z=1 and z=2 (an additional contribution due to the
Galactic column density has been fixed assuming $N_H^{Gal} =
1.3\times10^{21}$cm$^{-2}$, \citealt{dl90}).
Errors are at 68\% confidence level, $\chi^2_\nu$/(d.o.f.) is for the z=1
case.} \label{tab_spec}
\end{center}
\end{table*}

\begin{table*}
\begin{center}
\begin{tabular}{ccc}
\hline \\
Time after GRB
(s) & magnitude &  Band \\
\hline \\
164-214             &  $17.81^{+0.23} _{-0.19}$ & V \\
214-264             &  $18.19^{+0.25} _{-0.21}$ & V \\
311-321             &  $16.86^{+0.51} _{-0.35}$ & V \\
383.5-393.5         &  $17.14^{+0.66} _{-0.41}$ & V \\
457.5-462.5         &  $17.08^{+0.63} _{-0.40}$ & V \\
526.5-536.5         &  $16.97^{+0.59} _{-0.38}$ & V \\
597.5-607.5         &  $17.23^{+0.76} _{-0.44}$ & V \\
668.5-678.5         &  $17.82^{+1.57} _{-0.62}$ & V \\
740.5-750.5         &  $17.59^{+1.22} _{-0.56}$ & V \\
811.5-821.5         &  $19.09^{+\infty}    _{-0.86}$ & V \\
882.5-892.5         &  $17.09^{+0.72} _{-0.43}$ & V \\
954-964             &  $17.38^{+1.16} _{-0.55}$ & V \\
1210-1300           &  $18.50^{+0.92} _{-0.49}$ & V \\
11354-12254         &  $20.57^{+1.88} _{-0.65}$ & V \\
\hline
282.5-292.5         & $17.35^{+0.74} _{-0.44}$ & U\\
353.5-363.4         & $>20.25$                   & U\\
425-435                & $17.46^{+0.90} _{-0.48}$ & U\\
496.5-506.5         & $17.77^{+1.24} _{-0.57}$ & U\\
567.5-577.5         & $18.02^{+1.83} _{-0.65}$ & U\\
639-649                & $18.58^{+\infty}      _{-0.92}$ & U\\
710.5-720.5         & $17.55^{+1.02} _{-0.52}$ & U\\
781.5-791.5         & $18.65^{+\infty}      _{-0.98}$ & U\\
852.5-862            & $17.95^{+2.40}  _{-0.69}$ & U\\
924-934               & $18.97^{+1.44}  _{-0.60}$ & U\\
1001.5-1101.5    & $18.19^{+0.49}  _{-0.33}$ & U\\
6477.5-7140       & $20.02^{+1.24}  _{-0.56}$ & U\\
\hline
30500               & $20.08\pm0.4  $                    & R \\
51900               & $20.73\pm0.05$                    & R \\
\hline
\end{tabular}
\caption{ Log of GRB~050712 Optical observations. Values quoted in
this table have been corrected for Galactic extinction. U and V results
come from the {\it Swift} UVOT, while the two R band measurements
have been performed by Zeh et al.~2005 (Tautenburg) and Maiorano
et al.~2005 (Bologna Observatory).}
t
\label{tab_obs}
\end{center}
\end{table*}

\clearpage

\clearpage

\clearpage

\clearpage

\clearpage



\clearpage

\begin{thebibliography}{}

\bibitem[\protect\citeauthoryear{Barthelmy et al.}{2005}]{bar05} Barthelmy
S.D. et al. 2005, \apj, 559, 710

\bibitem[\protect\citeauthoryear{Berger et al.}{2003}]{be03}
Berger E. et al., 2003, Nature, 426, 154

\bibitem[\protect\citeauthoryear{Bloom et al.}{1999}]{bloo99}
Bloom J.S. et al., 1999, Nature, 401, 453

\bibitem[\protect\citeauthoryear{Burrows et al.}{2005a}]{bur05a}
Burrows D.N. et al., 2005a, Space Science Review, in press

\bibitem[\protect\citeauthoryear{Burrows et al.}{2005b}]{bur05b}
Burrows D.N., Romano P., Godet O. et al., 2005b, X-ray Universe 2005
proceedings, astro-ph~0511039

\bibitem[\protect\citeauthoryear{Campana et al.}{2005}]{cam05}
Campana S. et al. 2005, A\&A, submitted

\bibitem[\protect\citeauthoryear{De Pasquale et al.}{2005}]{dp05}
De Pasquale M., Piro L. et al., 2005, A\&A submitted (astro-ph~0507708)

\bibitem[\protect\citeauthoryear{Dickey \& Lockman}{1990}]{dl90}
Dickey J.M. \& Lockman F.J. 1990, ARAA, 28, 215

\bibitem[\protect\citeauthoryear{Esin \& Blandford}{2002}]{es00}
Esin, A.A. \& Blandford, R., 2000, \apjl, 534, L51

\bibitem[\protect\citeauthoryear{Falcone et al.}{2005a}]{fa05a}
Falcone A. et al. 2005a, GCN 3573

\bibitem[\protect\citeauthoryear{Falcone et al.}{2005b}]{fa05b}
Falcone A. et al. 2005a, ApJ submitted.

\bibitem[\protect\citeauthoryear{Frontera}{2004}]{fr04} Frontera F.,
Amati L., Lazzati D. et al., 2004, \apj, 614, 301

\bibitem[\protect\citeauthoryear{Galama et al.}{2000}]{gal00}
Galama, T.J., et al., 2000, \apj, 536, 185

\bibitem[\protect\citeauthoryear{Garnavich et al.}{2000}]{gar00}
Garnavich, P., Loeb, A. \& Stanek, K., 2000, \apjl, 544, L11

\bibitem[\protect\citeauthoryear{Gehrels et al.}{2005}]{ge05}
Gehrels N. et al., 2005, \apj, 621, 558

\bibitem[\protect\citeauthoryear{Gendre et al.}{2005}]{gen05}
Gendre B. et al., 2005, A\&A submitted (astro-ph/0507710)

\bibitem[\protect\citeauthoryear{Grupe et al.}{2005a}]{gr05a}
Grupe D. et al., 2005a, GCN 3573

\bibitem[\protect\citeauthoryear{Grupe et al.}{2005b}]{gr05b}
Grupe D. et al., 2005b, GCN 3579

\bibitem[\protect\citeauthoryear{Grupe et al.}{2005c}]{gr05c}
Grupe D. et al., 2005c, to be submitted to \apjl

\bibitem[\protect\citeauthoryear{Guidorzi et al.}{2003}]{gu03}
Guidorzi C., Frontera F., Montanari E. et al. 2003, \apj, 401, 491

\bibitem[\protect\citeauthoryear{Hill et al.}{2005}]{hi05}
Hill J.E. et al., 2005, in preparation

\bibitem[\protect\citeauthoryear{Hjorth et al.}{2003}]{hjo03}
Hjorth J., Sollerman, J., M\o ller, P., et al. 2003, Nature, 423,
847

\bibitem[\protect\citeauthoryear{Huang et al.}{2004}]{hu04}
Huang Y.F., Wu X.F., Dai Z. G., et al., 2004, ApJ, 605, 300

\bibitem[\protect\citeauthoryear{Kumar \& Piran}{2000}]{kp00}
Kumar P. \& Piran A., 2000, ApJ, 541, L9

\bibitem[\protect\citeauthoryear{Kumar \& Panaitescu}{2000}]{zap00}
Zhang B. \& Panaitescu P., 2000, \apj, 541, L51

\bibitem[\protect\citeauthoryear{Krimm}{2005}]{krm05}
Krimm H.A. 2005, private communication

\bibitem[\protect\citeauthoryear{Lamb}{2003}]{lam03}
Lamb D.Q., 2003, proc. "Gamma-Ray Burst and Afterglow Astronomy
2001" Woods Hole, Massachusetts Journal-ref: AIP Conf.Proc. 662,
433, astro-ph/0210434

\bibitem[\protect\citeauthoryear{Lamb \& Reichart}{2000}]{lam00}
Lamb D.Q. \& Reichart D.E., 2000, ApJ, 536, 1L

\bibitem[\protect\citeauthoryear{Lazzati \& Perna}{2002}]{lap02}
Lazzati D. \& Perna R., 2002, MNRAS, 330, 383.

\bibitem[\protect\citeauthoryear{Markwardt et al}{2005}]{mar05}
Markwardt G. et al., 2005, GCN 3576

\bibitem[\protect\citeauthoryear{Maiorano et al.}{2005}]{mai05}
Maiorano E. et al., 2005, GCN 3601

\bibitem[\protect\citeauthoryear{M\'{e}sz\'{a}ros~ \& Rees}
{1997}]{mr97}
M\'{e}sz\'{a}ros~ P. \& Rees M.J. 1997, \apj, 476, 232

\bibitem[\protect\citeauthoryear{Nousek et al.}{2005}]{no05}
Nousek J. et al., to be submitted to \apj (astro-ph 0508332)

\bibitem[\protect\citeauthoryear{O'Brien et al.}{2005}]{obr05}
O' Brien P. et al., 2005, \apj submitted, (astro-ph/0601125)

\bibitem[\protect\citeauthoryear{Panaitescu et al.}{1998}]{pan98}
Panaitescu A., M\'esz\'aros, P. \& Rees, M.J. 1998, \apj, 503, 314

\bibitem[\protect\citeauthoryear{Panaitescu et al.}{2005}]{pan05}
Panaitescu A. et al. 2005, MNRAS submitted (astro-ph/0508340)

\bibitem[\protect\citeauthoryear{Preece et al.}{2000}]{pre00}
Preece R. D. et al., 2000, \apjs, 126, 19

\bibitem[\protect\citeauthoryear{Piro et al.}{2005}]{pi05}
Piro L., De Pasquale M., Soffitta P. et al., 2005, \apj, 623, 314

\bibitem[\protect\citeauthoryear{Protassov et al.}{2002}]{pro02}
Protassov R., van Dyk D.A, Connors A. et al. 2002, \apj, 571, 545

\bibitem[\protect\citeauthoryear{Reichart}{1999}]{rei99}
Reichart E., 1999, \apjl, 521, L111

\bibitem[\protect\citeauthoryear{Rol et al.}{2005}]{rol05} Rol
E. et al. 2005, GCN 3575

\bibitem[\protect\citeauthoryear{Romano et al.}{2005}]{romano05}
Romano  et al., 2005, in preparation

\bibitem[\protect\citeauthoryear{Roming et al.}{2005}]{rom05a}
Roming P. et al., 2005, Space Science Review in press

\bibitem[\protect\citeauthoryear{Sari \& Piran}{1999}]{sap99}
Sari R. \& Piran T., 1999, \apjl, 519, 17 (SP99)

\bibitem[\protect\citeauthoryear{Stanek et al.}{2003}]{sta03}
Stanek, K.Z., Matheson, T., Garnavich, P. M., et al., 2003, \apj,
591, L17

\bibitem[\protect\citeauthoryear{Zeh et al.}{2005}]{zeh04}
Zeh A., Klose S. \& Hartmann D.H., 2004, \apj, 609, 952

\bibitem[\protect\citeauthoryear{Zeh et al.}{2005}]{zeh05}
Zeh A. et al., 2005, GCN 3646

\bibitem[\protect\citeauthoryear{Zhang \& M\'{e}sz\'{a}ros~}{2001}]{zm01}
Zhang B. \& M\'{e}sz\'{a}ros~ P., 2001, \apj, 552L, 35

\bibitem[\protect\citeauthoryear{Zhang \& M\'{e}sz\'{a}ros~}{2002a}]{zm02a}
Zhang B. \& M\'{e}sz\'{a}ros~ P., 2002a, \apj, 581, 1236

\bibitem[\protect\citeauthoryear{Zhang \& M\'{e}sz\'{a}ros~}{2002b}]{zm02b}
Zhang B. \& M\'{e}sz\'{a}ros~ P., 2002b, \apj, 566, 712

\bibitem[\protect\citeauthoryear{Zhang et al.}{2003}]{zh03}
Zhang B., Kobayashi S. \&  M\'{e}sz\'{a}ros 2003, \apj, 595, 950

\bibitem[\protect\citeauthoryear{Zhang \& M\'{e}sz\'{a}ros~}{2004}]{zhm04}
Zhang B. \& M\'{e}sz\'{a}ros~ P., 2004, IJMPA, 19, 2385
(astro-ph~03111321)

\bibitem[\protect\citeauthoryear{Zhang et al.}{2006}]{zh05}
Zhang B. Fan Y.Z., Dyks J. et al 2006, \apj submitted
(astro-ph~0508321)

\bibitem[\protect\citeauthoryear{Zuo \& Lu}{1993}]{zuo93} Zuo L. \&
Lu L., 1993, \apj 418, 601

\end{thebibliography}
\end{document}